\documentclass[submission,copyright,creativecommons]{eptcs}
\providecommand{\event}{GCM 2025} 

\usepackage{iftex}

\ifpdf
  \usepackage{underscore}         
  \usepackage[T1]{fontenc}        
\else
  \usepackage{breakurl}           
\fi

\title{Introducing The Maximum Common Bigraph Problem}
\author{Kyle Burns \qquad Michele Sevegnani \qquad Ciaran McCreesh \qquad James Trimble
\institute{School of Computer Science\\
University of Glasgow\\
Glasgow, UK}
\email{kyleburnsgla@gmail.com \qquad\qquad michele.sevegnani@glasgow.ac.uk}
\email{ciaran.mccreesh@glasgow.ac.uk \qquad\qquad james.trimble@yahoo.co.uk}
}
\def\titlerunning{Introducing The Maximum Common Bigraph Problem}
\def\authorrunning{Burns et al.}

\usepackage{graphicx} 

\usepackage{graphics}
\usepackage{url}
\usepackage{algorithmic}
\usepackage{comment}
\usepackage{algorithm}
\usepackage{mcaption}
\usepackage[font=scriptsize]{subfig}
\usepackage{amsmath}
\usepackage{amssymb}
\usepackage{bookmark}
\usepackage{multicol}
\usepackage[symbol]{footmisc}
\usepackage{amsthm}

\usepackage{color}
\usepackage{changebar}
\usepackage{soul}

\newcommand{\ifc}[2]{\langle#1,#2\rangle}
\newcommand{\df}{\stackrel{\text{\tiny def}}{=}}

\newcommand{\Nam}{\mathcal{X}}
\newcommand{\Sig}{\mathcal{K}}
\newcommand{\Ver}{\mathcal{V}}
\newcommand{\Edg}{\mathcal{E}}

\begin{document}

\theoremstyle{definition}
\newtheorem{definition}{Definition}[section]
\newtheorem{prop}{Proposition}

\maketitle

\begin{abstract}

Bigraph reactive systems offer a powerful and flexible mathematical framework for modelling both spatial and non-spatial relationships between agents, with practical applications in domains such as smart technologies, networks, sensor systems, and biology. While bigraphs theoretically support the identification of bisimilar agents, by simulating and comparing their corresponding minimal contextual transition systems, no known algorithm exists for computing the maximum shared structure between two bigraphs, an essential prerequisite for determining the set of possible transitions for a given agent state. In this work, we provide a definition of the maximum common bigraph problem, and present an adaptation of the McSplit maximum common induced subgraph algorithm to compute the maximum common bigraph between two bigraph states. Our approach opens a path toward supporting bisimulation checking in bigraph-based tools, which have been leveraged in other modelling paradigms for simplification, optimisation, and verification of models. 

\end{abstract}

\section{Introduction}

Bigraph reactive systems (BRSs), first introduced by Milner~\cite{Milner09} are a universal mathematical formalism which is capable of representing both spatial and non-spatial relations between entities, through the use of the two-tier bigraph data structure. Given a BRS model, consisting of an initial agent $G$ and set of reaction (rewriting) rules in the form $R \rightarrow R'$, a bigraph toolkit such as \textit{BigraphER}~\cite{BigraphER:2016} is able to perform a simulation of how the agent can potentially evolve over time by computing the raw \textit{transition system} of the model. The core component to this process is an efficient underlying matching algorithm, which can find all occurrences of $R$ in $G$ in order to substitute $R'$ at each step in a process known as \textit{bigraph matching}. Practical applications of BRSs include the simulation of processes in fields such as sensors in IoT devices~\cite{ModellingAndVerificationOfWSN:2018}, network communications \cite{80211:2014}, biological phenomena \cite{StochasticBigraphs:2008}, security for smart buildings~\cite{AutomatedMgmtOfSecurityIncidents:2019} and swarm programming for drones~\cite{UAVSwarmsBehavior2021,grzelak2025drone}. 

In addition to raw transition systems, BRSs in theory can also compute the \textit{minimal contextual} labeled transition system (MCTS) of a model. This process requires the finding of the largest structural overlap between a reaction rule redex $R$ and an agent state $G$ rather than a complete match, such that a wider environment can provide the missing minimal context to allow a full match (and subsequent substitution) to occur. This effectively models how an agent might evolve in \textit{any} context, rather than in a vacuum. Bisimulation, a property which can be utilised for the optimisation and verification of models, can be verified for bigraph agents by determining whether the MCTSs of two models are equivalent, as this guarantees that their behaviors will be identical when put into any possible context. Existing tools can presently only model raw transition systems, which are not powerful enough to guarantee this property. 

In this work, we provide a definition of the \textit{maximum common bigraph} (MCB) algorithm, which describes how to find the largest overlapping between two bigraph states, the prerequisite step necessary for finding all minimal contexts between a given agent and rule. We provide an encoding from bigraphs to graphs such that MCB can be solved using the \textit{McSplit} MCIS algorithm \cite{ijcai2017p99}, based upon earlier work which demonstrated that passing a similar encoding to a subgraph isomorphism solver resulted in very efficient solve times for full bigraph matching \cite{matching}.

\paragraph{Acknowledgment.} An expanded version of this work has been documented as part of the following Ph.D. Dissertation: \textit{Efficient and Scalable Algorithms for Bigraph Matching}, Kyle Burns, Glasgow University \cite{thesis}.

\section{Bigraphs}

\begin{figure}
    \centering
    \includegraphics[scale=0.35]{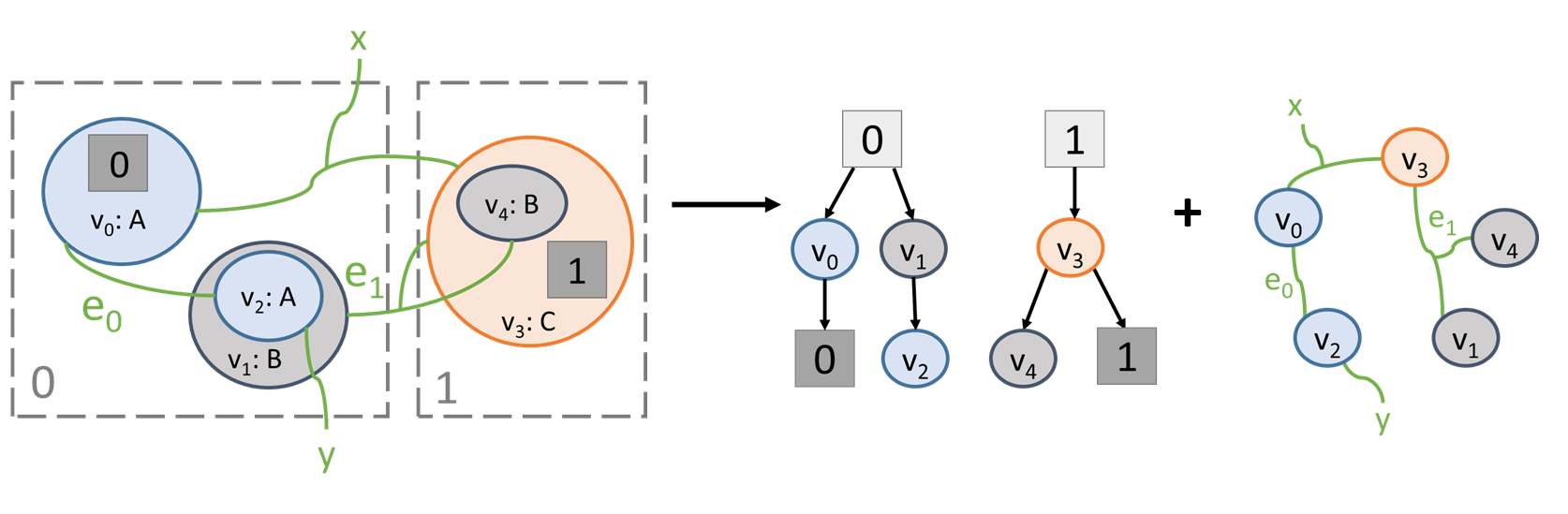}
    \caption{Example of a concrete bigraph with entities, their controls and their spatial and non-spatial relationships, and how the bigraph can be separated into its place and link graph components.}
\end{figure}

A bigraph $G$ is a graph-like data structure containing two separate components which share the same set of entities (nodes): the \textit{place graph}, a directed forest which represents the spatial relationships between entities (e.g.\ a device inside a room), and the \textit{link graph}, a hypergraph which represents non-spatial relationships (i.e.\ devices connected in a local network). While bigraphs can be represented both algebraically and graphically, we use graphical notation where possible in this work. 

A graphical example of a bigraph and its component place and link graphs is provided in Figure~1. Place graph relations are depicted as the parent encapsulating its child node(s), whereas link graph adjacencies are depicted using green links. All entities are also labelled (typed) with a \textit{control} value, represented graphically by color in the provided example, and each control has an associated \textit{arity} integer value which defines how many \textit{ports} (link graph adjacencies) each entity with that control must have. The place graph has $n$ \textit{regions} and $m$ \textit{sites}, top and bottom level places respectively, which are shown diagrammatically as numbered squares. These represent abstractions of unknown (or empty) additional structures above and beneath the current bigraph state, and denote where the composition of another place graph can be allowed to take place. Similarly, unconnected links extending above and below the current bigraph are known as its set of \textit{outer names} $Y$ and \textit{inner names} $X$, and denote where a composition of another link graph is available. A link is \textit{open} if it connects to an inner or outer name, and \textit{closed} if it only connects between entities. 

The set of $m$ sites and inner names $X$ make up the \textit{inner face} of a bigraph, denoted as $\langle m, X \rangle$. Conversely, the set of $n$ regions and outer names $Y$ make up its \textit{outer face}, denoted as $\langle n, Y \rangle$. Taken together, this describes the \textit{interface} of a bigraph, which is written as $G: \langle m, X \rangle \rightarrow \langle n, Y\rangle$. Sites and regions are identified using an ordered set of non-negative integers, e.g.\ $n = 3$ indicates that the set of regions is $\{0, 1, 2\}$. The interface of the bigraph provided in Figure 1 can thus be deduced as $G: \langle 2, \{y\} \rangle \rightarrow \langle 2, \{x\} \rangle$. The example bigraph provided is also a \textit{concrete} bigraph, meaning that all entities $\{v_0$, ..., $v_4\}$ and closed links $\{e_0, e_1\}$ in addition to the interface components are assigned a unique identifier. It is also a \textit{solid} bigraph, meaning that there exists no direct adjacency between the inner and outer faces (e.g. an edge connecting a region and site), no two inner face components are siblings, and each outer face component has at least one adjacency. For applying a rewrite rule $R \rightarrow R'$ to a bigraph $G$, both $R$ and $G$ are always solid.

Through the interfaces of bigraphs, a larger concrete bigraph $G = A \circ B$ can be built from smaller component concrete bigraphs $A$ and $B$ through connecting the inner face of $A$ with the outer face of $B$, and connecting on like-names. An alternative form of composition is also the tensor product $G = A \otimes B$ of two bigraphs, which simply places the two structures side by side. These form the basis for the occurrence, decomposition and rewriting rules of bigraph reactive systems. We now provide the formal algebraic definition of a bigraph in order to introduce BRSs and types of transition systems. To supplement our definition, we define the set of names used to identify entities, controls and closed links as belonging to the disjoint infinite sets $\Ver, \Nam$ and $\Edg$ respectively.

\subsection{Bigraph Definitions}

\begin{definition}[Concrete Place Graph]
    A \textit{concrete place graph} $B= (V_B,ctrl_B,prnt_B):m \rightarrow n$ is a triple which has the inner face $m$ and outer face $n$, indicating $m$ sites and $n$ regions. $B$ has a finite set $V_B \subset \Ver$ of entities, a control map $ctrl_B: V_B \rightarrow \Sig$, and a \textit{parent map} $prnt_B: m \uplus V_B \rightarrow V_B \uplus n$ that is acyclic i.e.\ $(v,v) \not\in prnt^+_B$ for any $v\in V_B$, with $prnt^+_B$ the transitive closure of $prnt$.
\end{definition}

\begin{definition}[Concrete link graph]
A \emph{concrete link graph} $B = (V_B, E_B, ctrl_B, link_B): X \rightarrow Y$ is a quadruple having (finite) inner name set $X \subset \Nam$ and an outer name set $Y\subset \Nam$. $B$ has finite sets $V_B \subset \Ver$ of entities and $E_B \subset \Edg$ of links, a control map $ctrl_B: V_B \rightarrow \Sig$ and a \emph{link map} $link_B : X \uplus P_B \rightarrow E_B \uplus Y$ where $P_B \df \{(v,i) \mid v \in V_B,\  0 \le i < ar(ctrl_B(v)) \}$ is the set of ports of $B$, and $ar(ctrl_B(v))$ is the arity value of $v$'s control. \emph{Closed links} are those where the domain is restricted to $P_B$ and the image is in $E_{B}$---otherwise they are \emph{open}. In addition, \textit{idle edges} are links where the domain is restricted to $\emptyset$, i.e.\ have no source to point from. 
\end{definition}

\begin{definition}[Concrete bigraph] 
A concrete bigraph $B =(V_B, E_B,ctrl_B, prnt_B,link_B):\ifc{m}{X} \rightarrow \ifc{n}{Y}$ consists of a concrete place graph $B^{\sf P}=(V_B,ctrl_B,prnt_B): m \rightarrow n$ and a concrete link graph $B^{\sf L}=(V_B,E_B,ctrl_B,link_B):X \rightarrow Y$. The inner and outer \emph{interfaces} of $B$ are $\ifc{m}{X}$ and $\ifc{n}{Y}$, respectively. The support size $\vert B \vert $ of the concrete bigraph is $\vert V_B \uplus E_B \vert$. A concrete bigraph is the combination of a concrete place graph and concrete link graph that each share the same entity set $V_B$.

\end{definition}

\begin{definition}[Concrete occurrence]
    Given two concrete bigraphs $A$ and $B$, it is said that $B$ occurs in $A$ if there exists some \textit{context} bigraph $C$ and \textit{parameter} bigraph $D$ such that $A = C \circ (B \otimes id) \circ D,$ where $id$ is the identity bigraph---an entity-free bigraph instance which only contains abstractions and faces that directly connect from $\langle m, X \rangle$ to $\langle n, Y \rangle$, allowing for place graph edges and link graph hyperedges to connect directly between $C$ and $D$ where necessary to ensure interface compatibility.
\end{definition}

\subsection{BRS Transition Systems}

A BRS model consists of an initial bigraph agent $A$ and a set of reaction rules of the form $R \rightarrow R'$. When an occurrence of $R$ exists in $G$ such that $A = C \circ (R \otimes id) \circ D$ for some context and parameter bigraphs $C$ and $D$ respectively, this permits a change in state (e.g.\ a device moving from one room to another) conducted through the substitution of $R$ with $R'$ to produce the resultant agent $G' = C \circ (R' \otimes id) \circ D$. A key part of this process is the finding of all occurrences of $R$ within an agent, which is denoted as the \textit{bigraph matching} problem. When there exists a match, this can be represented as the mapping of all support elements (entities and closed links) in the pattern rule to a matching element in the agent. This mapping is also known as an \textit{embedding} of $R$ inside $A$ \cite{dirBigEmbeddings}. Bigraph matching is an NP-hard problem, and bigraph toolkits often rely on SAT solving \cite{sharing,michelePhd}, constraint solving \cite{dirBigEmbeddings,matching} tools to perform this task. Another known approach is to transform bigraphs into a graph format which can be parsed using existing graph transformation tools to perform the full rewriting process, through a combination of subgraph isomorphism and single \cite{BiGGer2024} \cite{10.1145/3736704} or double \cite{GASSARA201973} pushout construction.

\begin{figure}
    \centering
    \includegraphics[scale=.2]{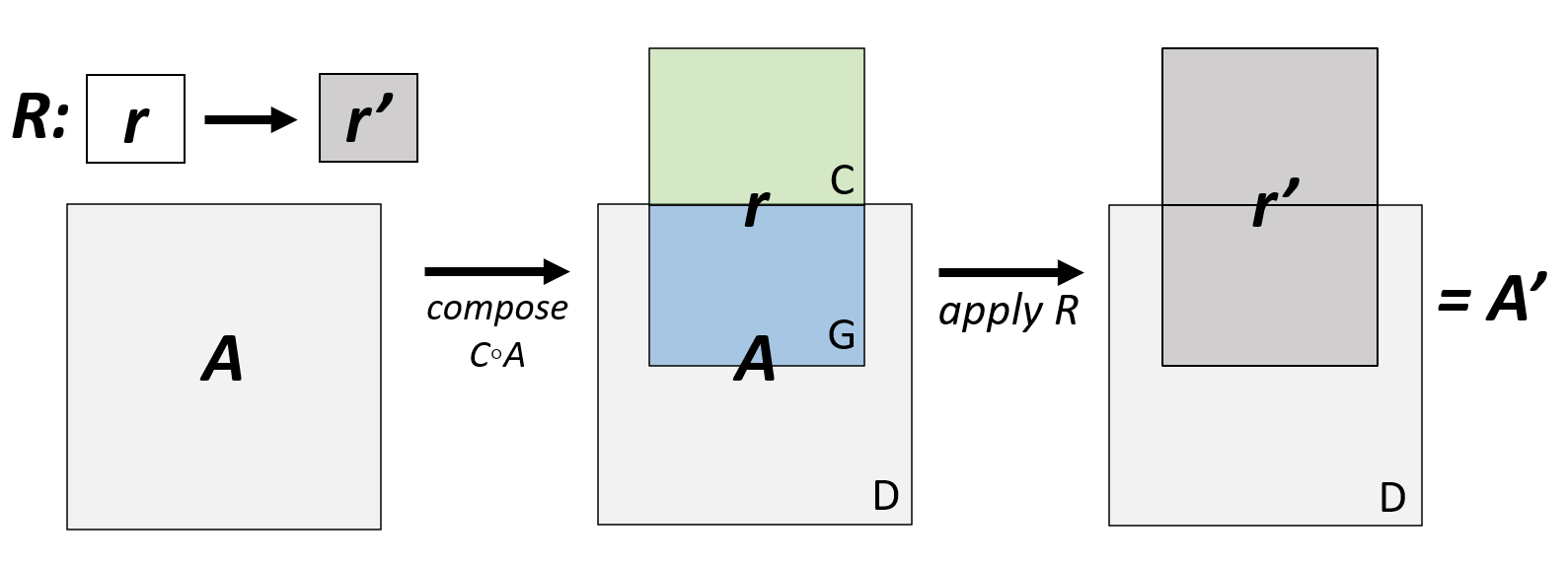}
    \caption{A high level representation of a transition in an LTS, where a context $C$ from the environment of agent $A$ is first composed to produce $C \circ A$, such that redex $r = C \circ G$ now occurs in the agent and the reaction rule $R: r \rightarrow r'$ can be applied. $G$ is the overlapping structure that occurs in both $r$ and $A$, and parameter bigraph $D$ is the non-overlapping remainder of $A$.}
\end{figure}

Through recursive exhaustive application and substitution of reaction rules upon an agent, this generates the \textit{raw transition system} of the model, a directed graph structure where nodes represent agent states which denotes all possible resultant states that can be reached, and edges represent the application of a reaction rule to reach one state from another. Raw transition systems are limited in how much they can model however, as they presume that the simulated agents must exist in a vacuum without any additional context, instead of as a component within a wider surrounding environment. In reality, there may be unknown entities and links connected to the agent we wish to account for, which can impact the behavior of how the state could potentially evolve over time. In a \textit{labelled} transition system (LTS), a transition is denoted as $A \xrightarrow[]{f} A'$, where $f$ is a context bigraph providing environmental structure. This context is composed onto the agent $A$ before applying a reaction rule, enabling transitions that depend on external structure. Raw transition systems can be considered a restricted form of LTSs where $f = \emptyset$, meaning reaction rules only apply when a full match already occurs. While LTSs offer a more expressive modelling framework by accounting for environmental influence, their generality introduces challenges: there are potentially infinite applicable contexts $f$, and the composition location within $A$ is underspecified. To address this, Milner \cite{Milner09} proposes a \textit{contextual transition system} (CTS), where transitions are annotated as $A \xrightarrow[]{(f, j)} A'$, and $j$ specifies the mapping of interface components where $f$ composes onto $A$. This can be further refined with the notion of a \textit{minimal transition}, where $f$ provides only the minimum amount of structure necessary for a reaction to occur. A \textit{minimal CTS}, or MCTS, includes only such transitions, which is sufficient for capturing any possible state the agent may reach through influence from its environment. An example application of a minimal context for an LTS is shown in Figure~2.

The primary motivation for implementing MCTS simulations is the ability to ensure \textit{bisimilarity} of agents in a model. A bisimulation between two agent states $A$ and $B$, denoted as $A \sim B$, is a symmetric relation such that both agents exhibit indistinguishable behavior under all transitions, and can be treated as functionally equivalent regardless of structural differences \cite{Milner09}. Bisimulation is a powerful model-checking tool with real-world applications in database refactoring \cite{bisim-database}, graph processing \cite{bisim-graph}, and reinforcement learning \cite{bisim-drl}, as it enables simplification, optimsation, and verification of system components. In the context of BRSs, this congruency can also be guaranteed through equivalence checking the respective MCTS tree structures of two agents, which can be performed in polynomial time. This could enable potential practical applications such as model simplification (substitution of complex components with smaller bisimilar states), behavioral verification (ensuring no unexpected behavioral changes after an update), and fault tolerance (components identified as bisimilar can be utilised as backups). 

\begin{figure}
    \centering
    \includegraphics[scale=.28]{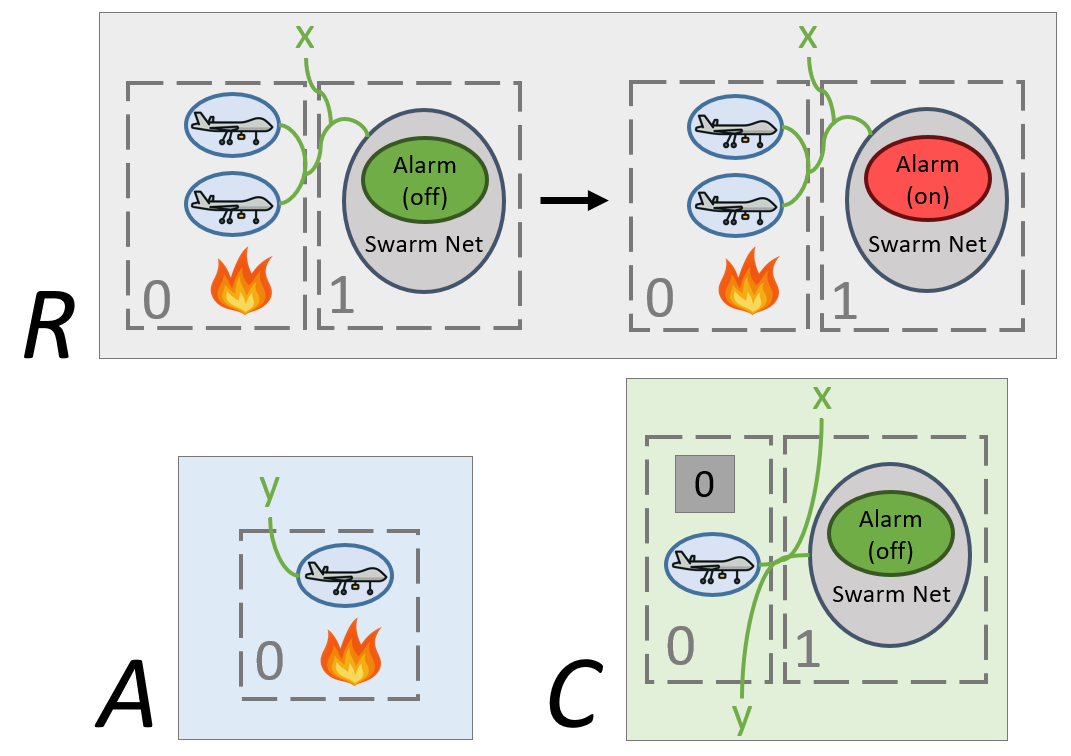}
    \caption{An example practical application of a MCTS transition when modelling drone swarms for wildfire detection. When context $C$ is supplied to the agent $A$, this allows reaction rule $R$ to apply.}
\end{figure}

A more practical example of MCTSs is provided in Figure 3, where a partial BRS model for simulating swarm behavior for drones is provided. The reaction rule $R$ defines the behavior that when two drones connected to the swarm network are positioned over a fire, this will activate the network's alarm. However, if we want to simulate all possible outcomes starting from a single drone discovering a fire as modelled by agent $A$, $R$ cannot be applied in a raw TS as it cannot take into account any wider environment (and thus any nearby drones) extending beyond $A$. A MCTS however is capable of modelling the potential of another drone and swarm network being in range, and can hence model this by supplying the minimal context $C$ to $A$ in order to account for this possibility.

In order to simulate a MCTS in practice, this requires a method for finding the minimal context $C_m$ for each permissible transition. This can be calculated through finding the largest overlap shared between a reaction rule $r$ and an agent state $A$ such that $r = C_m \circ G$ and $A = C_r \circ G \circ D_r$ where the resultant bigraph $C_m \circ A = (C_m \otimes C_r) \circ (r \otimes id) \circ D_r$ is produced. An example of this is shown in Figure 2 where it can be observed that $C_M$ can be obtained by first finding $G$, then decomposing all elements of $G$ from $R$ (for simplicity, $C_r$ is not graphically shown here). This follows a similar principle of using maximum common subgraph algorithms to solve graph edit distance problems \cite{BUNKE1997689}. 

\section{Maximum Common (Induced) Subgraph}

The \textit{maximum common subgraph} (MCS) problem is an NP-complete optimisation problem, which seeks to find a graph structure $G_S$ which exists as an isomorphism to a subgraph inside two given input graphs $G_1$ and $G_2$, where there does not exist any larger (by some metric) subgraph which meets the same criteria \cite{MCS}. There exist two main forms of MCS: maximum common \textit{induced} subgraph (MCIS), which is concerned with maximizing the number of vertices that can be mapped between $G_1$ and $G_2$, and the \textit{maximum common edge subgraph} problem (MCES) which instead uses the number of edges of $G_S$ as the measure of size to maximise. Going forward, we are primarily concerned with MCIS, as this more accurately reflects the behavior of bigraphs as described later. We formally define this as follows.

\begin{definition}[Maximum Common Induced Subgraph]
    Given two input graphs $G_1 = (V_1, E_1)$ and $G_2 = (V_2, E_2)$, a common induced subgraph $G_S = (V_S, E_S)$ is one such that there exists a pair of injective mappings $f_1: V_S \rightarrow V_1$ and $f_2: V_S \rightarrow V_2$, where vertices and edges and mapped such that $(u, v) \in E_S \iff (f_1(u), f_1(v)) \in E_1$ and $(u, v) \in E_S \iff (f_2(u), f_2(v)) \in E_2$.

    $G_S$ is \textit{maximum} when there exists no alternate solution $G_S' = (V_S', E_S')$ which satisfies the above conditions and also satisfies $ \vert V_S' \vert > \vert V_S \vert$.
\end{definition}

The current state of the art for MCIS is the \textit{McSplit} branch and bound algorithm, which models the problem similarly to a constraint solver by assigning vertices from the graph $V_{1}$ to $V_{2}$, backtracking when no more valid assignments are possible \cite{ijcai2017p99}. McSplit utilises \textit{label classes}, groups of unassigned vertex pairs that share identical adjacency patterns to already matched vertices, which are refined at each assignment step based on neighbourhood compatibility. These classes are efficiently maintained using three arrays: two permutations of vertex ids ($V_1$ and $V_2$) and a label class descriptor array $LC$, allowing all refinements and backtracks to be executed in $O(|V_1| + |V_2|)$ time without storing explicit bitsets. At each state, McSplit computes a tight upper bound on the number of further assignments possible as follows: $$bound = |M| + \sum\limits_{l \in LC} min(|(u \in V_1 \land LC(u) = l) |, |(v \in V_2 \land LC(v) = l)|)$$

If the bound cannot exceed the current best solution, the branch is pruned. There also exist optimised implementations of McSplit with efficient heuristic strategies based upon the PageRank algorithm \cite{icsoft23}. McSplit has additionally been proven to be extensible for supporting further features such as vertex labels/types, directed edges, enumerating all solutions and enforcing connectedness, which are later shown to be necessary to support the encodings of bigraphs \cite{ijcai2017p99}.

\section{Maximum Common Bigraph}

We now formally introduce our notion of the maximum common bigraph (MCB) problem as follows.

\begin{definition}[Maximum Common Bigraph]~\label{def:mcb}
    Given two input solid bigraphs $G_1$ and $G_2$, a member of the set of \textit{maximum common bigraphs} $(G_M, M) \in \texttt{MCB}(G_1, G_2)$ is a pair containing the solid bigraph $G_M$ which satisfies
    $$G_1 = C_1 \circ (id \otimes G_M) \circ D_1, \  G_2 = C_2 \circ (id \otimes G_M) \circ D_2$$ 
    such that there does not exist some other bigraph $G'_{M'}$ which satisfies
    $$\displaylines{
        G_1 = C_3 \circ (id \otimes G'_{M'}) \circ D_3, \ G_2 = C_4 \circ (id \otimes G'_{M'}) \circ D_4,\cr
        (\vert G'_{M'} \vert > \vert G_M \vert) \ \lor \ (G'_{M'} = C_5 \circ G_M \circ D_5, \ C_5 \neq id, \ D_5 \neq id)\cr
    }$$
    
    for some arbitrary context and parameter bigraphs $C_{1-5}$ and $D_{1-5}$ respectively.
    
    $M = \{(m_1, m_2, ..., m_n)\}$ is a set of triples $m_k = \{(g_{k}, a_{k}, b_{k})\}$, representing the embedded mapping between the entities, ports and closed links of $g \in G$ to those in $a \in G_1$ and $b \in G_2$ respectively. We wish to enumerate all such instances and their corresponding mappings.
\end{definition}

\begin{figure}
    \centering
    \includegraphics[scale=.2]{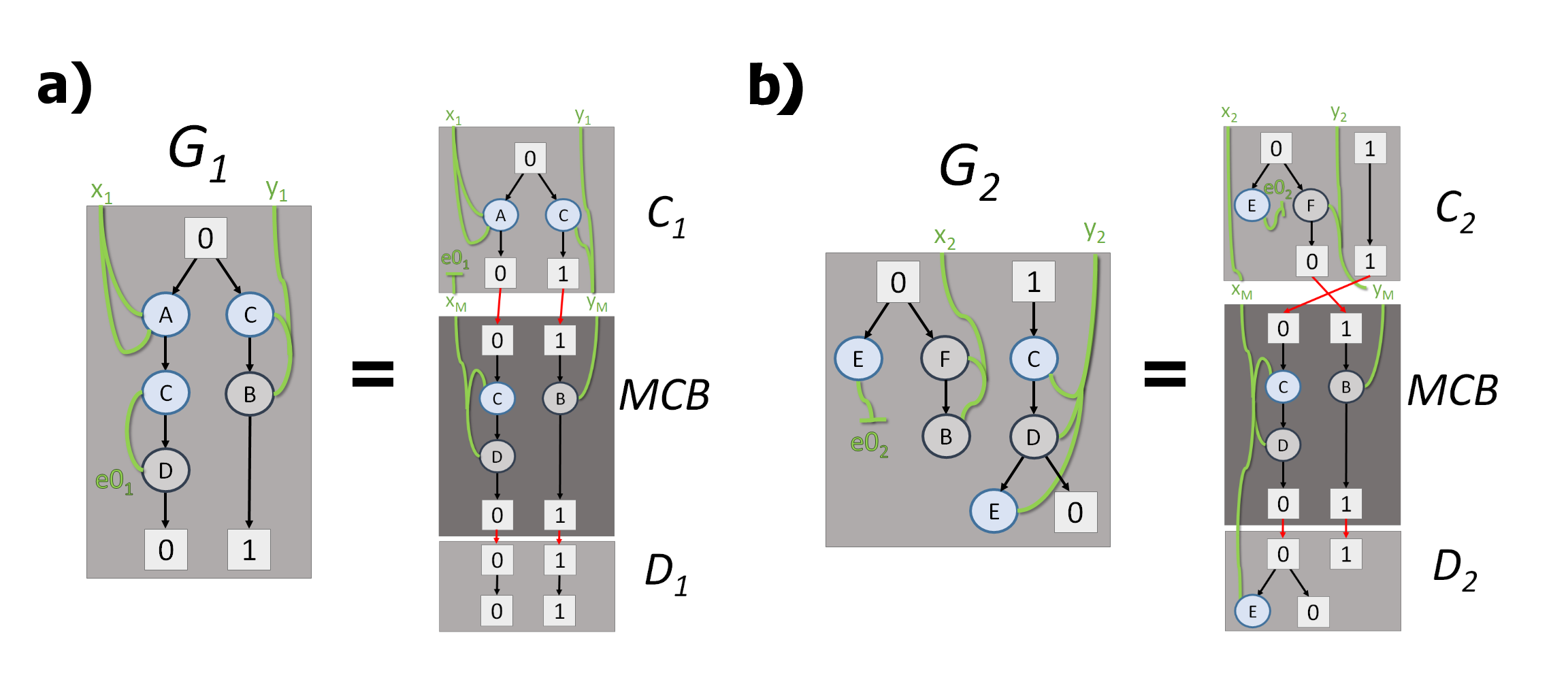}
    \caption{An instance of the maximum common bigraph problem, with $G_1$ and $G_2$ as the input bigraphs and $MCB$ as the solution. (a) A decomposition of $G_1$ to show that $MCB$ exists as a component in the bigraph. (b) A similar decomposition of $G_2$ to show $MCB$ also exists. The ports of $C$ and $D$ must be joined in $MCB$ as this still allows for a valid solution.}
\end{figure}

It can be observed that from this definition, a candidate solution $G_M$ must meet two separate requirements to be considered a maximum:

\begin{enumerate}
    \item There cannot exist any other solution bigraph $G'_M$ with a greater support size.
    \item There cannot exist any other solution bigraph $G_M'$ where $G_M$ occurs in $G_M'$ and is not isomorphic.
\end{enumerate}

The first requirement is clear as any candidate solution violating this cannot reasonably be considered the largest shared region between $G_1$ and $G_2$. The second requirement enforces that any solution must also meet the criterion of being maximal in the compositional sense---even if the first condition is met, the second can still be violated if any interfaces of $G_M$ can be merged or closed through further composition while remaining a sub-component of both $G_1$ and $G_2$. This extra step is necessary in order to restrict a theoretically infinite number of additional variants for each solution bigraph containing extra arbitrary regions/sites/links, as well as ensuring that $G_M$ still adheres to the conceptual notion of a maximal bigraph---we wish to ``saturate'' each discovered solution such that any further non-identity composition onto $G_M$ at all would no longer cause it to be a valid solution. This is handled through computing the \textit{relative pushout} (RPO) of the triple relation $(G_1, G_2, G_M)$ upon determining the optimal embedding. As the bigraph literature already describes how to calculate RPOs when given a pair of bigraphs and a mapping of elements between them \cite{Milner09}, we do not repeat this here, and instead focus on leveraging this approach to compute the required mapping as a prerequisite step. Figure 4 demonstrates an instance of the MCB problem, and the corresponding decompositions of $G_1$ and $G_2$ to retrieve the common bigraph.

Similarly to how MCIS can be considered an optimisation problem variant of the subgraph isomorphism (SIP) decision problem, we define MCB as the optimisation variant of bigraph matching. We also propose that MCB can be solved by treating the problem instance as MCIS with additional constraints and pre/post processing to handle the complexities introduced by hyperedges and compositional rules, as has been done in previous work to encode bigraph matching instances as SIP problems \cite{matching}.

\subsection{Properties}

Through this definition, we can intuitively infer some properties of MCB which must always hold for any instance.

\begin{itemize}
    \item \textbf{Identity}: $(G, M) \in \texttt{MCB}(G, G)$ for any bigraph $G$. 
    The maximum common bigraph between two equivalent bigraphs will also be an equivalence. There can also exist additional solution mappings when symmetries exist within $G$.
    \item \textbf{Inverse Rule}: $(G_M, M) \in \texttt{MCB}(G_1, G_2) \longleftrightarrow (G_M, M^{-1}) \in \texttt{MCB}(G_2, G_1)$ for any bigraph pair $G_1$ and $G_2$. Swapping the order of input graphs should still produce the same set of solutions, but with all mappings between $G_1$ and $G_2$ inverted.
    \item \textbf{Succession}: The solution $(G_M, M) \in \texttt{MCB}(G_3, \ \texttt{MCB}(G_1, G_2))$ represents the largest shared area between three bigraphs $G_{1-3}$, regardless of order of operations (transitive).
    \item \textbf{Matching}: $(S_{big} \neq \emptyset) \in \texttt{MATCH}(P, T)  \longrightarrow \ (P, S_{big}) \in \texttt{MCB}(P, T)$--if a full match of pattern bigraph $P$ exists in the target $T$, it follows that $P$ will also be a maximum common bigraph between the two with the same embedding.
\end{itemize}

\section{Encoding McSplit to Solve MCB}

We now go on to describe how McSplit can be adapted to support bigraphs encoded as directed/labelled standard graphs. We assume that instances are non-trivial, i.e.\ must contain at least one entity and no idle edges/regions/sites. The full proof of soundness and completeness of our approach is provided in Appendix A.

Firstly, we consider only the encoding of place graphs in isolation, and the changes we make to the McSplit array structures which enable us to enforce valid solutions which adhere to bigraphical composition rules. We then introduce our link graph flattening function to model the ports of entities and their relations, and the further adaptations made to the refinement process and reward function to support these. For each of places and links, their respective RPOs are then computed to ensure compositional saturation. A high level view of this process is demonstrated visually in Figure 5.

\begin{figure}
    \centering
    \includegraphics[scale=.25]{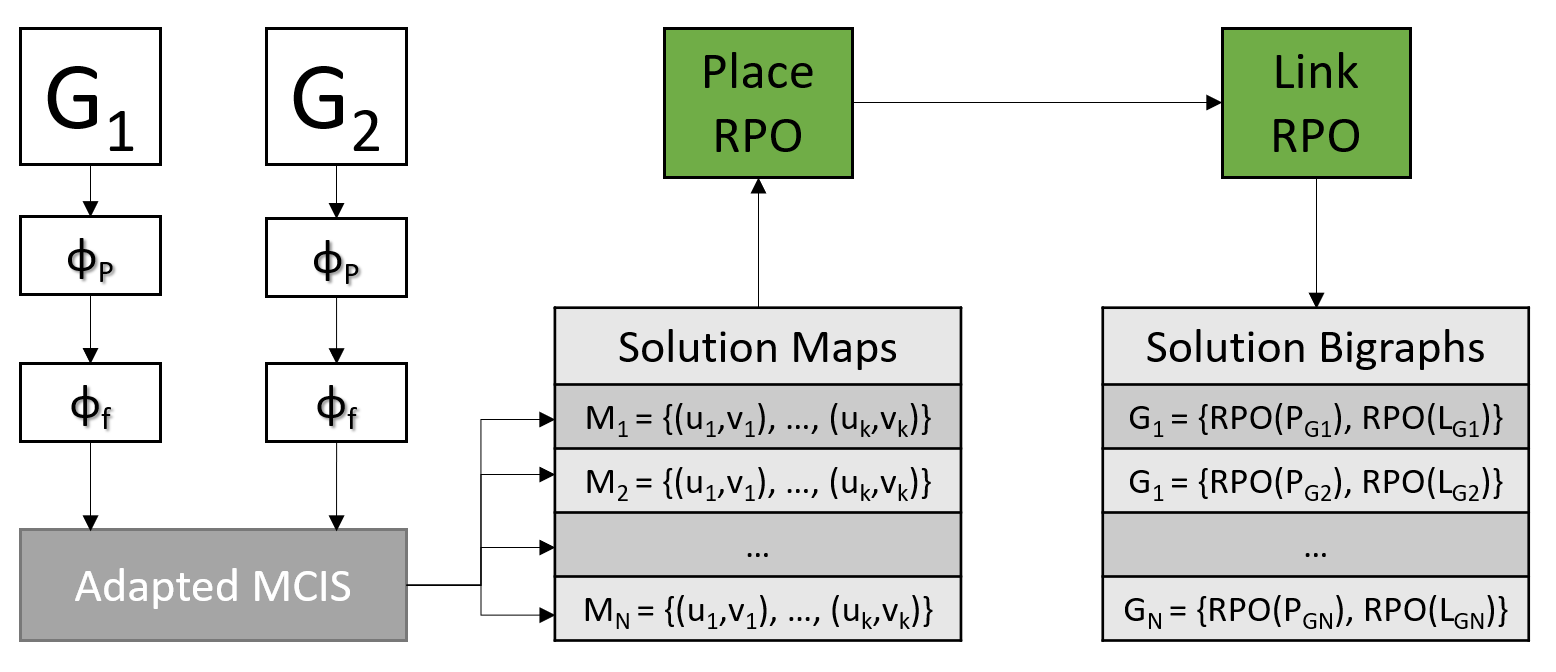}
    \caption{The MCB solving process, where the largest common areas between two encoded bigraphs are found using a MCIS algorithm before their interfaces are refined to produce a set of minimally bound solution bigraphs.}
\end{figure}

\subsection{Place Graph Encoding}

The place graphs of both $G_1$ and $G_2$ are encoded through discarding sites and regions as well as any parent relations to and from them respectively. We are not immediately concerned with these site and region placements of $G_1$ and $G_2$ when performing the underlying MCIS search loop, because a site/region placement (or lack of thereof) cannot ever cause an otherwise valid set of entity mappings for a candidate MCB to be invalidated; in the worst case, we can simply assign a site as a child to all entities $v \in V_{G_M}$ and a region as a parent to every top-level entity in the building of the solution bigraph(s), in order to guarantee that the result will always occur in $G_1$ and $G_2$ where it is found in its encoded form. Hence, we first assume as a baseline that this relaxed structure --- where all entities are adjacent to both the inner and outer interface --- will be the form of the returned solution bigraph(s) upon completing MCIS search, and allow the RPO function to refine the interface of the solution by closing and merging all regions and sites where possible. Controls are also encoded as node labels, to enforce the corresponding bigraph constraint that like must match with like.

After applying the encoding function $\phi_P$, the encoded versions of $G_1 = (V_{G_1},ctrl_{G_1},prnt_{G_1}): i \rightarrow j$ and $G_2 = (V_{G_2},ctrl_{G_2},prnt_{G_2}): k \rightarrow l$ take the form of the following graph pair: $$\phi_P (G_1) = \{V_{G_1}, (u, v) \in prnt_{G_1}^{-1} \ | \ v \neq i, u \neq j \}$$ $$\phi_P (G_2) = \{V_{G_2}, (u, v) \in prnt_{G_2}^{-1} \ | \ v \neq k, u \neq l \}$$

\subsubsection{Ensuring Valid Compositions}

The next step to implementing MCB for the place graph is modifying McSplit to ensure that a solution bigraph $G_M$ will always respect the compositional property of bigraphs, and thus exist as a component in $G_1$ and $G_2$ in the forms $G_1 = C_1 \circ G_M \circ D_1$ and $G_2 = C_2 \circ G_M \circ D_2$ respectively. To ensure this, all pairs of assigned entities must either be directly adjacent to one another (i.e.\ producing a connected subgraph), or fully transitively disjoint where neither are descendants of the other (i.e.\ producing a tensor product). We begin by considering McSplit for maximum common \textit{connected} induced subgraph, which enforces connected solutions through only selecting vertices from label classes if that class has at least one adjacency to the current solution subgraph. However, we want to relax this restriction in a way which still allows disjoint vertices to be selected, as long as they cannot be transitively reached by or from any current assigned vertex. In a CSP format, this would be added to the list of constraints as follows: $$\{\forall u, v \in G_1 \ |\ \{\forall v' \in prnt^{-1}_{G_1}(v) \ | \ \texttt{match}(v') = \emptyset\}\land  (u, v) \in prnt^+_{G_1} \land \texttt{match}(u) \neq \emptyset\} \rightarrow \texttt{match}(v) = \emptyset$$ Where $\texttt{match}(u) = v \leftrightarrow (u, v) \in M$. The same constraint applies to $G_2$, with the difference that all calls to the $\texttt{match}$ function are replaced with $\texttt{match}^{-1}$ (where $\texttt{match}^{-1}(u) = v \leftrightarrow (v, u) \in M$) to ensure symmetry.

This approach also requires that we initially construct two descendant maps for all pairs of vertices within each of $G_1$ and $G_2$ prior to beginning search --- which we define as $\tau_1$ and $\tau_2$ respectively --- where $\tau_{\{1, 2\}}(u, v)$ is true if and only if entities $u$ and $v$ are transitively adjacent. This allows the solver to determine whether an entity belonging to a ``no adjacencies'' label class should still be allowed to be selected for assignment. 

\begin{figure}
    \centering
    \includegraphics[scale=.35]{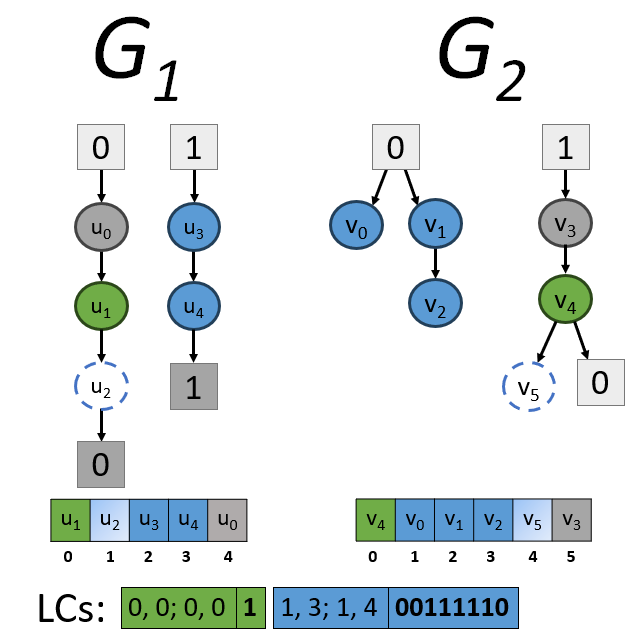}
    \caption{The adapted McSplit algorithm for a pair of place graphs after the mapping $u_0 \rightarrow v_3$ is made. Entities $u_2$ and $v_5$ are restricted from selection at the next assignment step, as they are transitively but not directly adjacent to $u_0$ and $v_3$ respectively. The blue label class vertices, which would be unselectable in default connected McSplit, are allowed to be selected here as they can exist as a tensor product of $u_1$ and $v_4$ respectively. For simplicity, controls are not shown in this example.}
\end{figure}

\begin{algorithm}[ht]
    \caption{Refine LC Visibility (int u, int v)}
    \label{alg:LC}
    \begin{algorithmic}
        \FORALL {$l \in LC$}
            \STATE $\mathbf{S_1} \leftarrow \{w \in V_{G_1} \land lc(w) = l\}$
            \STATE $\mathbf{S_2} \leftarrow \{w \in V_{G_2} \land lc(w) = l\}$
            \IF{$A'(l)[0] = 1 \ \OR \ (|S_1| > 0 \ \AND \ A_{V_1}[S_1[0]][u] > 0) \ \OR \ (|S_2| > 0 \ \AND \ A_{V_2}[S_2[0]][v] > 0)$}
                \STATE{$A(l) \leftarrow 1$}
                \STATE{\textbf{continue}}
            \ENDIF
            \STATE{$A(l)[0] \leftarrow 0$}
            \IF{$\ell(l) \neq \textbf{link}$}
                \FORALL{$u' \in S_1$}
                    \IF{$A'(l)[u'+1] = 1 \ \AND \ R[u][u'] = 1$}
                        \STATE{$A(l)[u'+1] \leftarrow 0$}
                    \ENDIF
                \ENDFOR
                \FORALL{$v' \in S_1$}
                    \IF{$A'(l)[v'+1] = 1 \ \AND \ R[v][v'] = 1$}
                        \STATE{$A(l)[v'+1] \leftarrow 0$}
                    \ENDIF
                \ENDFOR
            \ENDIF
       \ENDFOR
       
    \end{algorithmic}
\end{algorithm}

Connected McSplit can be adjusted to accommodate this at the variable selection step by replacing the boolean flag within each label class (signifying whether its nodes are adjacent to a candidate solution node) with a bitset variable which we define as $A(l)$. This indicates entity selection visibility for its member entities: $A(l)$ simply contains a 1 (true) value if the label class $l$ is adjacent to the current solution bigraph as it can be inferred that all member vertices will also be adjacent. However, if the label class is disjoint from the solution bigraph, then the bitset will begin with a zero followed by $(|\{u \in G_1 \land LC(u) = l\} | + |\{v \in G_2 \land LC(v) = l\})$ bits, where $A(l)[k+1]$ indicates whether the $k$th entity in the class is a current valid selection. During variable selection, a vertex $v$ is prohibited from being assigned to if $A(LC(v))[0] = 0$ and $A(LC(v))[v+1] = 0$, indicating that it is currently invisible to the propagator. Figure 6 provides an example instance of this adapted approach for a given state, where each LC is shown alongside their $A(l)$ values. For simplicity, this conditional visibility constraint does not apply to the flattened ports of an entity for this approach (Section 5.2), and thus an entity's port vertices can only be assigned once the entity itself has been assigned. 

We also define $A'(l)$ as the value of the label class $l$'s bitset at the previous assignment step. Algorithm 1 demonstrates the label class bitset refinement process upon making the assignment $(u \in V_{G_1} \rightarrow v \in V_{G_2})$, which is performed prior to the partitioning process. It can be observed that our bitset structure can be refined at an upper bound of $O(n + m)$ time similarly to the main McSplit label class refinement process, by checking each unassigned entities' relation to the entity in each bigraph assigned at that step. The partitioning process itself is also modified such that each entity's corresponding label class bit value is repositioned alongside them when appropriate, to maintain congruency between states.

Our encoding function, in addition to this additional refinement process to enforce compositional integrity, is sufficient to find all valid assignments between the maximum common bigraph(s) between the entities of a pair of place graphs. The discarded regions and sites of the graphs are later re-added and then closed/merged where appropriate as part of the post-process RPO function to guarantee maximality.

\subsection{Link Graph Encoding}

We now consider the addition of the link graph into our encoding. We proceed by constructing an appropriate link flattening function.

Given a bigraph $B^{\mathsf{L}}: (V_B, E_B, ctrl_B, link_B): X \rightarrow Y$, and the encoding of its place graph $D$ $\phi_{\{P,T\}}(D^\mathsf{P}) : (V_{D}, E_{D})$, where (where $V_B=V_D$), we define the flattening function as follows:
$$\phi_{f} : \phi_{\{P,T\}}(D^\mathsf{P}) \times B^{\mathsf{L}} \mapsto (V, E)$$

The vertices of the resultant flattened graph can be described as: $$ V = V_{D} \uplus P_{B} \uplus \widehat{E_B}, \quad \widehat{E_B} = \{ e \in E_B \mid link_B(p) = e, p \notin X \}$$

where $\widehat{E_B}$ is the set of closed links in $B^{\mathsf{L}}$, $P_B$ are the ports of $B^{\mathsf{L}}$, and an additional \textit{closure} node is added to represent each closed link. We re-use the bigraph \emph{edge} identifier as a vertex identifier in the flattened graph. 

We describe the resultant edge set as follows: $$E = E_D \uplus \{(v,p) \mid p = (v,i) \in P_B \} \uplus \{(p,e) \mid e \in \widehat{E_B}, link_B(p) = e\}$$

\begin{figure}
    \centering
    \includegraphics[scale=.25]{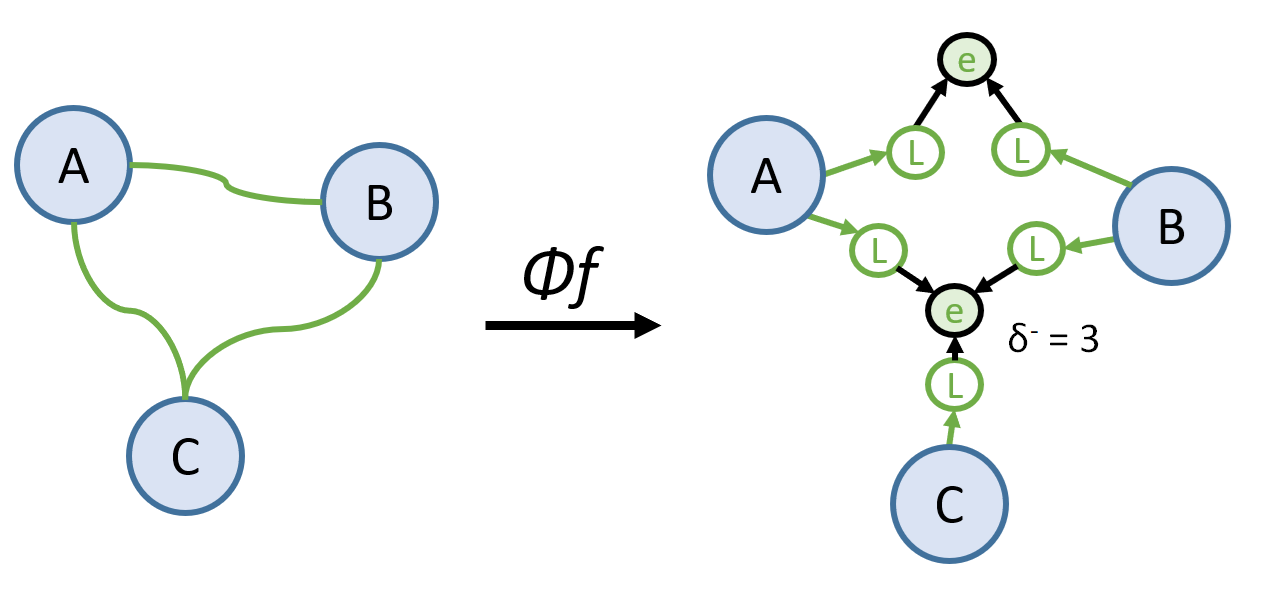}
    \caption{An example of link flattening for the MCB algorithm --- hyperedge connections are not considered for the main search loop, and are instead re-added and merged upon finding a solution.}
\end{figure}

This is a less constrained variant of the clique-based link flattening function used for bigraph matching \cite{matching}, where we no longer build cliques between ports which share a hyperedge. This allows the flattened graph to represent a version of the input link graph pair where all ports have been freed up and each have a lone connection to the bigraph interface, which is the link graph's equivalent to our process of discarding all sites and regions in the place graph. This produces a link graph where there are no constraints between interface components, representing the minimal possible structure that can be a substructure of $G_1$ and $G_2$ when a valid set of mappings are found. Similarly to the place graph, then later refine and merge all possible sets of links (retaining $G_M$ as a sub-bigraph of $G_1$ and $G_2$) post-search. A visual application of the modified flattening function is provided in Figure 7.

An additional degree equality constraint is placed on closure nodes to ensure that any common hyperedge will have isomorphic adjacency sets in $G_1$ and $G_2$. In McSplit, this can be preliminarily enforced using the existing label class structures by further partitioning the label class representing all $\ell = \textbf{closure}$ nodes by in-degree prior to search. To preserve the structure of the solution bigraph's hyperedges, we also constrain closure nodes such that they are only available for variable selection once \textit{all} of their parent port nodes have been assigned. This prevents the case where a solution may contain a ``closed'' link edge that is still connected to the interface, which we seek to prohibit. This can be reflected by enforcing the additional constraint: $$\{\forall e \in E_{G_1}, \ p = (v, i) \in P_{G_1} \ | \ (p, e) \in link_{G_1} \land \ \texttt{match}(p) = \emptyset\} \rightarrow \texttt{match}(p) = \emptyset$$

This can be implemented by checking neighbouring closures upon port assignment and toggling them to visible if all their parent ports have been assigned, using the existing bitset structure in the label class to distinguish between visible and invisible closure vertices. An additional adjustment to achieve this is that closure label class bitsets will always have their first bit set to zero, as adjacency to the currently selected substructure alone no longer guarantees visibility for selection. Enforcing this on $G_1$, in combination with the degree constraint to ensure like-like matching, is sufficient to also ensure this holds for closures in $G_2$ as sharing a label class will guarantee that their respective neighbourhoods will match.

\subsubsection{Modifying the Reward Function}

This method of representing ports as flattened vertices introduces a discrepancy between the size of a candidate solution's encoded form and its support size. An example of where this arises is shown in Figure 8, which shows a MCB instance in the form of a pair of bigraphs, and its corresponding MCIS instance after encoding and flattening $G_1$ and $G_2$. The MCB instance has two solutions of equal size 1, that is, $(A \rightarrow A)$ and $(B \rightarrow B)$. However, in the encoded MCIS instance, only the $(B \rightarrow B)$ solution will be returned, because its ports will also contribute toward the size of the mapping, meaning that the MCIS algorithm will consider the mapping of $(B \rightarrow B)$ and its ports a solution of size 3, compared to mapping $(A \rightarrow A)$ and its lone port to produce a solution of size 2. Hence, we wish to modify McSplit to ignore the assignment of port vertices, specifically when enumerating the current score of a candidate solution. However, an occurrence of an assigned entity with an unassigned port can never appear in a valid solution, as all components of the entity must be matched alongside the entity itself (as matching labels always have the same arity value, all ports should always be available for matching). Taking this into account, we introduce a new scoring function to determine the optimality of a solution as follows:

\begin{equation*}
    score = 
    \begin{cases}
        -1 \ \ \  \text{if } \exists \{ \ p = (v, i) \in P_{G_1} \ | \  \texttt{match}(p) = \emptyset \ \land \texttt{match}(v) \neq \emptyset \}\\
        |G_M| \ \ \ \text{otherwise}
    \end{cases}
\end{equation*}

\begin{figure}
    \centering
    \includegraphics[scale=.22]{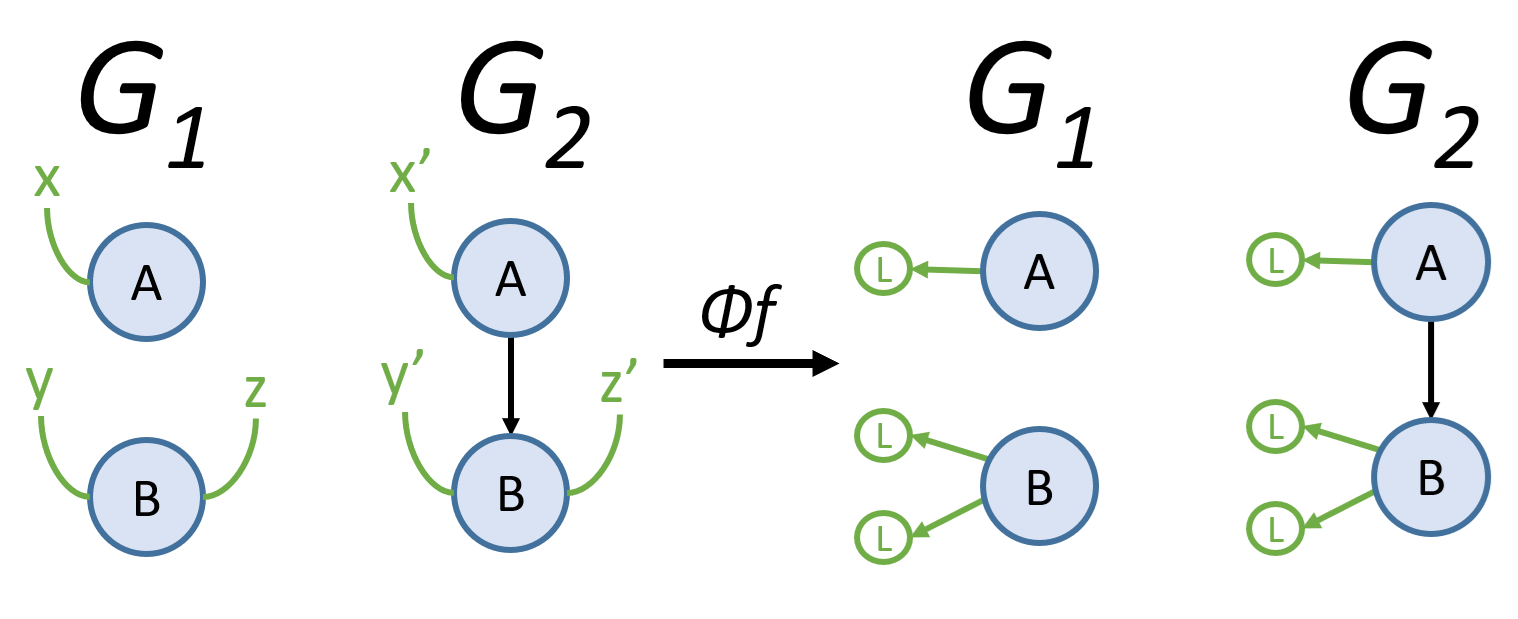}
    \caption{An example of a flattened MCB instance where the returned MCIS solution set does not match the expected solution set of MCB due to counting port vertices, without first modifying the score function.}
\end{figure}

Where $P_{G_1}$ is the set of flattened ports of bigraph $G_1$, and $|G_M|$ is the support size of the common bigraph. The symmetry of assigned entities and arities means only one of the input bigraphs needs to be checked to verify validity. In addition, we also modify the McSplit bound function to ignore all label classes containing port nodes as follows, since they do not contribute to the score of a solution:

$$bound = |G_M| + \sum\limits_{l \in LC} min(|(u \in \{V_{G_1} \uplus E_{G_1}\} \land LC(u') = l \ \land \ u \neq (p, i) \in P_{G_1}) |,$$

$$|(v \in \{V_{G_2} \uplus E_{G_2}\} \land LC(v') = l \ \land \ v \neq (q, j) \in P_{G_2})|)$$

Where $u'$ and $v'$ are the encoded vertices representing support elements $u$ and $v$. As vertex label types are partitioned into separate label classes during initialization, it is trivial to determine whether a label class $l$ is a ``port'' class by checking the vertex label value of any member element, and skip $l$ during summation if the label function of the vertex $\ell(v) = \textbf{link}$. While restricting the upper bound is not a necessary step to ensure correctness of solutions, it is in our interest to do so whenever possible to ensure optimal performance and minimise time spent performing redundant search tree traversal, as long as no valid solution is ever incorrectly filtered as a result.

\section{Implementation}

We demonstrate a prototype solver for our adapted MCB algorithm, which was implemented by modifying a variant of McSplit provided by Trimble, written in Python and originally created as a contribution to the \textit{NetworkX} Python graph library \cite{networkx}. This variant implements a simplified and unoptimised version of McSplit which does not make use of the efficient label class structure described in Section 3, but instead simulates their behavior through creating new Python list structures which store the label class membership of each vertex at each refinement step. Whilst not as efficient as an optimised implementation of McSplit engineered using a more efficient language, experimentation with this simplified implementation allowed for easier engineering and prototyping of the necessary constraints to model MCB, which could then be evaluated for correctness using a mix of manually crafted instances and a suite of bigraph matching instances later used for benchmarking. 

This implementation also makes use of the McSplit$\downarrow$ version of the algorithm, where the solver treats the instance as a sequence of decision problems. Beginning with $n = min(|G_1|, |G_2|)$, the search loop attempts to find a maximum common subgraph of size $n$, then iteratively decrements $n$ by 1 when an UNSAT is returned until at least one solution is found. The bound function is also modified to trigger a backtrack when the bound is strictly less than the goal, rather than less than or equal. It was found that McSplit$\downarrow$ performs narrowly better overall than default McSplit on a variety of evaluated benchmark instances \cite{trimble2023partitioning}.

We make our code publicly available \cite{mcb}. During experimentation, it was found that restricting the variable selection heuristic to only allow assignments to ports for $n$ steps after assigning to an entity of arity $n$ in order to enforce the ``no entities can have unassigned ports'' constraint resulted in improved solve times. This is achieved through keeping track of a global $port\_lock$ variable, which is set to $n$ whenever selecting an entity of arity $n$, and then decremented by 1 every time a port vertex is selected. Whenever $port\_lock > 0$, the score of the current set of assignments is set to zero and considered an invalid solution. Because of the connected-ness constraint on ports, an entity's port vertices only become visible for selection once the vertex itself has been selected, ensuring that this logic is sufficient to guide the variable selection order appropriately. 

\subsection{Evaluation}

We provide a preliminary evaluation of our prototype tool, using a selection of 2000 generated bigraph matching instances based upon Milner's conference call model example \cite{Milner09}. Since MCTSs take a reaction rule redex and agent state as input similarly to bigraph matching, and both search for occurrences of the pattern in order to perform a rewrite and expand their respective transition systems, these test instances are suitable for use in this context. The pattern rules (producing encoded graphs of size $|V| = 6$) are matched against target agents of varying sizes, ranging from encoded graph sizes of 59 up to 1581. We record the time elapsed upon calling the underlying MCIS search function on the encoded graphs ignoring input/output time, and also include the time taken to build each solution mapping's corresponding RPO bigraph after search. We ensure the confidence of our MCB implementation by verifying that for each instance with at least one full occurrence in bigraph matching, that each of its occurrences also exist in the solution set of MCB --- a required property which we identify in Section~4.1. We also verify the symmetry of MCB by ensuring that when the order of the input graphs are reversed, that the same set of solutions (with inverted mappings) are always returned for the full suite of instances. We make the full raw dataset available \cite{burnsdata}.

\begin{figure}
    \centering
    \includegraphics[scale=.4]{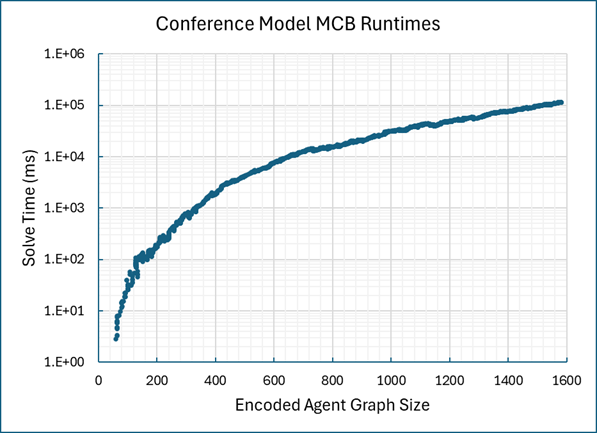}
    \caption{Comparing the solve time of each conference model redex/agent pair as a function of the encoded agent graph sizes (ranging from 59 to 1581) across a test suite of 2000 instances.}
\end{figure}

Figure 9 shows the solve times achieved by our prototype tool compared against the encoded graph size of the agent. From this comparison, we can determine that our tool was able to solve encoded target states with a graph size of up to roughly 300 within one second, showing promise for being able to solve bisimulations for small to moderately sized agents using this implementation. Out of the 2000 instances, our prototype was able to solve roughly 400 within one second, further demonstrating feasibility for computing the MCTS of smaller scale models. While this includes the additional time spent building the RPO bigraphs of each solution, this was found to take $<1\%$ of the total solve time across all instances, and thus was found to be trivial in this context. This was primarily due to the small size of the reaction rule redexes, although in a practical context these pattern rules would be expected to be small regardless (e.g.\ the largest average pattern match size across all real-world example BRSs provided by the BigraphER suite was found to be 9). However, we note that it may be of interest to record the impact on performance as the provided pattern bigraph increases in size, up to (or even greater than) the size of the target, as part of a full evaluation for a future optimised solver. Notably, bisimulations for bigraphs cannot be verified for models which produce infinite transition systems, as this signals that there are infinite possible states that each agent can evolve into --- at most, a model will only be able to determine ``bisimilar up to N steps'' for a pair of agents in this case. Thus, it would not be expected that the agents grow to absurdly large sizes in practice.

\section{Conclusion}

In this paper, we have provided a definition for the maximum common bigraph problem, which we identify as an optimisation-based generalization of bigraph matching where solutions adhere to the RPO property of bigraphs in order to guarantee maximality on compositions. We also describe how this can be used to find the inverse of the smallest possible composition that allows a reaction rule to be applicable, which is the key problem that must be solved within minimal contextual transition systems, and identify additional constraints to guarantee interface compatibility between the context and agent. We then build a prototype backtracking solver based upon the McSplit algorithm to solve MCB. While this implementation is presently an unoptimised prototype, its evaluated performance on a selection of rule/agent input pairs found that it is theoretically feasible for use on smaller scale models. 

A clear avenue is presented for further research --- now that we have introduced the problem and its use cases, we now wish to build upon this approach and adapt the McSplit MCB algorithm to handle the additional constraints imposed when finding all minimal contexts for a agent/rule pair in practice. For example, a MCB solution may not necessarily provide a valid MCTS if there is a mismatch on the interface which blocks the missing context from being composed, and this could ideally be detected and excluded from consideration during search propagation. A more optimised low-level solver, capable of using bit-parallel structures and exploiting McSplit's full capabilities, could then be integrated into a bigraph toolkit like BigraphER, where further evaluation via building MCTSs for a suite of BRSs modelling real-world use cases could be performed.

This approach also shows potential to be improved upon through parallelization. While efficient subgraph isomorphism solvers are able to efficiently handle SIP instances containing thousands of vertices, state of the art MCIS algorithms can only computationally handle up to approximately 35 vertices on unlabeled graphs before becoming impractical in the general case \cite{MCSSIP}, and thus we can expect MCB (and subsequently MCTS building) to be more computationally expensive than performing bigraph matching in a raw TS. This motivates the idea of solving multiple MCB instances for a MCTS simultaneously where possible by sending them to separate process threads --- if a dedicated MCB solver were integrated into a toolkit such as BigraphER, support for parallel transition system growth could be integrated in a solver-agnostic manner without being concerned what type of transition system (raw or minimal contextual) is being built or what solver is being relied upon under the hood.

\addcontentsline{toc}{chapter}{Bibliography}
\bibliographystyle{eptcs} 
\bibliography{dissertation}
\newpage

\appendix

\section{Proof of Soundness and Completeness}

Here we provide a proof of soundness and completeness of our adaptation. Soundness is proven by demonstrating that any solution identified by the MCIS model corresponds to an instance of a solution which adheres to our definition of a maximum common bigraph. Conversely, completeness is proven by showing that when a solution exists in an instance of MCB, the adapted algorithm will also find a corresponding match in the MCIS encoding.

As it is already established in the literature of bigraphs how to construct an RPO for a pair of bigraphs and a valid set of entity/closure mappings between them \cite{Milner09}, we do not repeat the proof of this here --- instead, it is satisfactory to prove that the set of assignments returned by the adapted McSplit solver is sufficient for then passing to the RPO building functions to retrieve the maximum common structure. With that in mind, we represent each solution $M$ returned by modified McSplit as an injective set of $(u, v)$ pairs denoting the mapping from each vertex $u$ in $G_1$ to $v$ in $G_2$, ignoring the vertices of the later constructed common bigraph (handled by the RPO functions). Thus we wish to prove that for any maximum common bigraph embedding of size $m$: $$S_{MCB} = \{(u_1, v_1), ..., (u_m, v_m)\}$$ The corresponding encoding will produce a solution: $$M = \{(u_1', v_1'), ..., (u_m', v_m')\}$$ with a bijective relation between $(u_k, v_k)$ and $(u_k', v_k')$ and vice versa. We propose some observations related to bigraph composition in order to supplement our proofs.

\begin{prop}

Given the composition of two place graphs $$G: m \rightarrow n = (A: k \rightarrow n)\circ (B: m \rightarrow k)$$ if $u \in V_B, v \in V_G$ and $u \in prnt^{+}_G(v) $ then $v \in V_B$. 

\end{prop}

This states that any descendant of an entity in $V_B$ must also be in $V_B$. We prove this through the recursive application of the definition of bigraph composition \cite{Milner09}, where for some $v' \in V_G$, if $prnt_G(v') \in V_B$ then $v' \in o \uplus V_B$, and we know $v'$ is an entity and therefore $v' \in V_B$. Applying this to $u$ means that $prnt^{-1}(u) \subseteq V_B$, and this can be recursively applied to all children of $u$ to prove that all grandchildren of $u$ exist in $V_B$ and so on, and this can be repeated recursively in order to reach all descendants of $u$. Hence, for any $v$ such that $prnt^+_G(v) = u \in V_B$, we show that $v \in V_B$ as required.

\begin{prop}

Given the composition of two place graphs $$G: m \rightarrow n = (A: k \rightarrow n)\circ (B: m \rightarrow k)$$ if $u \in V_A, v \in V_G$ and $v \in prnt^{+}_G(u) $ then $v \in V_A$. 

\end{prop}

This states that any ancestor of an entity in $V_A$ must also be in $V_A$. We again prove this through the recursive application of the definition of bigraph composition \cite{Milner09}, where for some $v' \in V_A$, $prnt_G(v') = prnt_A(v')$ and hence $prnt_G(v') \in V_A$. Applying this to $prnt_G(u)$ means that the grandparent of $u$ will be in $V_A$, and this can be recursively applied at each upper depth to prove that all ancestors of $u$ must exist in $V_A$. Hence, for any $v$ such that $prnt^+_G(u \in V_A) = v$, we show that $v \in V_A$ as required.

\begin{prop}

Given the composition of three place graphs $$G: m \rightarrow n = (A: k \rightarrow n)\circ (B: l \rightarrow k)\circ (C: m \rightarrow l)$$ if $u \in V_B, v \in V_B, w \in V_G$ and $w \in prnt^{+}_G(u), v \in prnt^+_G(w)$ then $w \in V_B$. 

\end{prop}

This states that any entity that exists between a pair of entities in $V_B$ must also be in $V_B$. We prove this by applying Propositions 13 and 14 together as follows: by 1, as $u$ is an ancestor of $w$ and $u \in (B \circ C)$, then we know $w \in (B \circ C)$. By Proposition 2, as $v$ is a descendant of $w$ and $v \in (A \circ B)$, then we know $w \in (A \circ B)$. Taken together, we can conclude that $w$ can only exist in $B$, and therefore $w \in V_B$. As a corollary, we can informally deduce that for any two entities $(u \in V_G, v \in V_G)$, if $u \in V_B$ and $v \in V_B$ and $prnt_G^+(u) = v$, then \textit{all} entities between them must also exist in $B$.

\begin{prop}

Given the composition of two link graphs $$G: X \rightarrow Y = (A: Z \rightarrow Y)\circ (B: X \rightarrow Z)$$ for any $p \in P_B$ and $e \in E_B$, $link_G(p) = e$ if and only if $link_B(p) = e$.

\end{prop}

This states that a port in $B$ is linked to an edge if and only if they are also linked in $G$. Firstly, from the definition of bigraph composition \cite{Milner09}, $link_B(p) = e \in E_B \implies link_G(p) = link_B(p)$, therefore $link_G(p) = e$. Secondly, we prove that $link_B(p) = e \impliedby link_G(p) = e$ by hypothesizing a $p \in P_B$, $e \in E_B$ such that $link_G(p) = e$ but $link_B(p) \neq e$. This would instead mean $link_B(p) = x \in (Z \uplus E_B)/\{e\}$. If this were the case, then by link graph construction, either $link_G(p) = x \neq e$ if $x \in E_B$ which is a contradiction, or $link_G(p) = link_A(x)$ if $x \in Z$. However as $link_A(x)$ exists in $A$, then it cannot be $e \in E_B$ which contradicts our hypothesis and proves our proposition by contradiction. As a corollary, we deduce that $\vert link_G^{-1}(e) \cap P_G\vert = \vert link_B^{-1}(e) \cap P_B\vert$.

\subsection{Soundness}

Given an instance of MCB $(G_1, G_2)$ with a McSplit encoding $(\phi_{f}(\phi_{P}(G_1)), \phi_{f}(\phi_{P}(G_2)))$ for which there exists a solution of size $m$ in the form of the injective mapping $M = \{(u_1', v_1'), ..., (u_m', v_m')\}$, we wish to prove that this corresponds to a MCB solution $G_1 = C_1 \circ (G_M \otimes id) \circ D_1$, 
$G_2 = C_2 \circ (G_M \otimes id) \circ D_2$, in the form of an injective embedded mapping $S_{MCB} = \{(u_1, v_1), ..., (u_m, v_m)\}$ from a subset of support elements $u \in G_1$ to a support element $v \in G_2$, where $u_k' \rightarrow v_k'$ is the encoded form of support elements $u_k \rightarrow v_k$. We wish to prove two key attributes: firstly, that the bigraph $G_M$ constructed from $M$ is common to $G_1$ and $G_2$. Secondly, that there are no alternate set of assignments $M'$ of size $>m$ which is also common to $G_1$ and $G_2$.

\subsubsection{Commonality}

We begin our proof of commonality by constructing the input bigraphs $G_K , \ K = \{1, 2\}$ from their encoded forms. We perform this for the place graphs $P = (V_{K}, ctrl_{K}, prnt_K)$ as follows: \begin{align*}
& V_K =  V_{\phi_{P}(G_K)} \\
& \{\forall v_i \in V_K \ | \ ctrl_K(v_i) = \ell(v'_i)\} \\
& \{\forall (v_i, v_j) \in V_K \ | \ (v_i', v_j') \in E_{\phi_{P}(G_K)} \rightarrow prnt(v_j) = v_i \} \\
\end{align*} 

The link graphs $L = (V_K, E_K, ctrl_K, link_K)$ can then be constructed as follows: \begin{align*}
& V_K = \{ g' \in \phi_{f}(G_K) \ | \ \ell(g') \notin \{\mathbf{link}, \mathbf{closure}\}\} \\
& P_K = \{ g' \in \phi_{f}(G_K) \ | \ \ell(g') = \mathbf{link}\} \\
& E_K = \{ g' \in \phi_{f}(G_K) \ | \ \ell(g') = \mathbf{closure}\} \\
& \{\forall g \in V_K \ | \ ctrl_{K}(g) = \ell(g')\} \\
& \{\forall (g_a', g_b') \in E_{\phi_{f}(G_K)} \ | \ \ell(g_a') = \mathbf{link}, \ell(g_b') = \mathbf{closure}\} \rightarrow link_G(g_a \in P_G) = g_b \in E_G
\end{align*}

We note that the structural information regarding the interfaces of $G_1$ and $G_2$ are lost as part of the initial encoding process. However, as previously discussed, these interface components cannot impact whether or not the set of assignments $M$ produce a valid MCB since the interface of $G_M$ is initially assumed to be in a fully open state, where the RPO functions handle the constraining of interface components later on --- hence, they can only affect the structure of the common bigraph RPO, and therefore we do not require them as part of proving the soundness of $M$. Using $M$ and either of our reconstructed input bigraphs (we use $G_1$ for this proof), the common bigraph $G_M$ with the least constrained possible interface (fully open from above and below by assigning sites and regions where possible) can thus be built by disregarding all elements which are not part of the solution mapping (and thus cannot appear in $G_M$).

We perform this for the place graph $P = (V_{M}, ctrl_{M}, prnt_M)$ as follows: \begin{align*}
& V_M =  \{ u \in V_{G_1} \ | \  M(u) \neq \emptyset\} \\
& \{\forall v \in V_M \ | \ ctrl_M(v) = ctrl_{G_1}(v)\} \\
& \{\forall (v_i, v_j) \in V_M \ | \ prnt_{G_1}(v_j) = v_i \rightarrow prnt_M(v_j) = v_i \} \\
& \{\forall v \in V_M \ | \ prnt_{M}(v) \notin V_M \rightarrow prnt_M(v) = r \in n \} \\
& \{\forall v \in V_M \ | \ (s \in m) \in prnt^{-1}_M(v) \} \\
\end{align*} We also add a new outer name as a sink for each $g \in P_M$ where $link(g) \notin E_M$. The link graph $L = (V_M, E_M, ctrl_M, link_M)$ can then be constructed as follows:\begin{align*}
& V_M =  \{ u \in V_{G_1} \ | \  M(u') \neq \emptyset\} \\
& P_M = \{ p \in P_{G_1} \ | \ M(p') \neq \emptyset\} \\
& E_K = \{ e \in E_{G_1} \ | \ M(e') \neq \emptyset\} \\
& \{\forall v \in V_M \ | \ ctrl_{M}(v) = ctrl_{G_1}(v)\} \\
& \{\forall (g_a', g_b') \in E_{\phi_{f}(G_K)} \ | \ \ell(g_a') = \mathbf{link}, \ell(g_b') = \mathbf{closure} \rightarrow link_G(g_a \in P_G) = g_b \in E_G\} \\
& \{\forall p \in G_M \ | \ link_{M}(p) \notin E_{M} \rightarrow link_{M}(g) = y \in Y\}
\end{align*}  Thus, we are able to build the common bigraph $G_M$. In doing so, this provides us the required set of triple relations $(u \in G_1, v \in G_2, w \in G_M)$, where each $w$ corresponds to the mapping of common elements $(u, v) \in M$. 

From here, we effectively reduce the problem to a pair of bigraph matching instances, where $G_M$ must occur as a pattern in both $G_1$ and $G_2$ in order to demonstrate that $G_1 = C_1 \circ (G_M \otimes id) \circ D_1$ and $G_2 = C_2 \circ (G_M \otimes id) \circ G_2$. The context and parameter components $C_K$ and $D_K$, $K = \{1, 2\}$ for each input bigraph can hence be reconstructed as follows, beginning with the place graphs: \begin{align*}
& \{\forall g_i \in V_M \ | \ prnt_M(g_i) = r \in n\} \rightarrow prnt_{G_K}(g_i) \in V_{C_K}, prnt_{C_K}(r) = g_i \\
& \{\forall g_i \in V_M, g_j \in V_{G_K} \ | \ prnt_M^{-1}(g_i) \cap m = s, prnt_{G_K}(g_j) = g_i, g_j \notin M\} \rightarrow g_j \in V_{D_K}, prnt_{D_K}(g_j) = m \\
& \{\forall g_i \in V_{G_K} \ | \ prnt_{G_K}(g_i) \in V_{D_K} \} \rightarrow g_i \in V_{D_K}, prnt_{D_K}(g_i) = prnt_{G_K}(g_i) \\
& \{\forall g_i \in V_{G_K} \ | \ g_i \notin V_M, g_i \notin V_{D_K} \} \rightarrow g_i \in V_{C_K} \\
& \{\forall g_i \in V_{C_K}\} \rightarrow prnt_{C_K}(g_i) = prnt_{G_K}(g_i) 
\end{align*} As this place graph construction already assigns all entities in $G_K$ to either $C_K$, $G_M$ or $D_K$, we assume this has been already performed when building the links of $C_K$ and $D_K$ as follows: \begin{align*}
& \{\forall p \in P_M \ | \ link_M(p) = y \in Y\} \rightarrow link_{C_K}(y) = link_M(p) \\
& \{\forall p \in P_{D_K} \ | \ link_{G_K}(p) \notin D_K\} \rightarrow link_{D_J}(p) = x \in X, link_{id}(x) = x' \in Y, link_{C_K}(x') = link_M(p) \\
& \{\forall p \in P_{C_K} \} \rightarrow link_{C_K}(p) = link_{G_K}(p) \\
& \{\forall e \in E_{D_K} \} \rightarrow link^{-1}_{D_K}(e) = link^{-1}_{G_K}(e)
\end{align*} This hence builds the full $G_K = C_K \circ (G_M \otimes id) \circ D_K$ decomposition on the entities and relations of the place graph, and ports and edges of the link graph. As this now gives us our initial pair of MCB compositions and the common bigraph $G_M$ from its MCIS encoding and set of vertex assignments $M$, this concludes our proof by construction.

\subsubsection{Maximality}

Maximality can also be proven by construction through analysis of the modified scoring function as follows. The score of $M$ will either be -1 if the current set of assignments decode into an incomplete bigraph, or conversely if the solution is structurally sound, the score of $M$ will be $\sum\limits_{(u', v') \in M} \{\ell(u') \neq \textbf{link}\}$. In any MCB instance, the minimum possible score of any instance will always be 0, where $M = \{\}$ and $G_M = \emptyset$, therefore an incomplete bigraph will never be a solution and we can assume $score(M) >= 0$ is always true for a final result.

From this, we can infer that the score of the MCIS solution $M$ will always equal the support size of $G_M$ in the original MCB instance, as ports do not count toward support size, and this is reflected in our MCIS adaptation by enforcing that their corresponding flattened link nodes do not count toward the scoring function. Therefore we can conclude through this bijection that the maximum scoring subgraph(s) returned by MCIS must always correspond to the maximum common structure(s) $G_M$ between $G_1$ and $G_2$. This concludes our proof.

\subsection{Completeness}

Given an instance of MCB $(G_1, G_2)$ where $G_1 = C_1 \circ (G_M \otimes id) \circ D_1$ and $G_2 = C_2 \circ (G_M \otimes id) \circ D_2$, for solid bigraphs $G_1$ and $G_2$ and the shared bigraph $G_M$, and there exists a solution in the form of an injective embedded mapping of size $m$, $S_{MCB} = \{(u_1, v_1), ..., (u_m, v_m)\}$ from a subset of support elements $u \in G_1$ to a support element $v \in G_2$, we wish to prove that a parallel solution of the same size $M = \{(u_1', v_1'), ..., (u_n', v_n')\}$ exists in the modified MCIS instance $(\phi_{f}(\phi_{P}(G_1)), \phi_{f}\phi_{P}(G_2))$, where $u_k' \rightarrow v_k'$ is the encoded form of support elements $u_k \rightarrow v_k$. Similarly to our soundness proof, we wish to prove two key attributes: firstly, that no valid MCB solution will be incorrectly filtered by the encoding or constraints. Secondly, that no optimal MCB solution will be incorrectly pruned by the bound function.

\subsubsection{Validity}

Assume that there exists a valid pair of compositions $G_1 = C_1 \circ (G_M \otimes id) \circ D_1$ and $G_2 = C_2 \circ (G_M \otimes id) \circ D_2$, with a corresponding embedding of $G_M$ in both bigraphs $S_{MCB} = \{(u_1, v_1), ..., (u_m, v_m)\}$ of size $m$, where the corresponding MCIS solution $M = \{(u_1', v_1'), ..., (u_m', v_m')\}$ is not a valid solution. This suggests that at least one of our newly added constraints are being violated. 

We first consider our method of encoding the bigraphs. By construction, all parent relations between entities in $G_K, K = \{1, 2\}$ are preserved through the encoding of $prnt_{G_{K}}(v) = u$ as the edge $(u', v')$ in $E_K$, and thus conventional MCIS holds. Trivially, as controls are preserved through graph labelling, McSplit will split any incompatible entities apart before search, ensuring only compatible entities can be mapped to one another. This also sufficiently handles flattened edge nodes as for any $(u', v')$, $u \in E_K \iff v \in E_K$. From Proposition 4, we know that the degree and solidity constraints on closure nodes will never be violated for a valid bigraph composition, as $link_{G_M}(p \in P_{G_M}) = e \in E_{G_M}$ if and only if $link_{G_K}(p) = e$.

Finally, we now consider our connectedness constraint on entities. By Proposition 3, we have proven that for all pairs of entities $(u, v) \in G_M$ which are transitively adjacent, then all entities between $u$ and $v$ must also be in $G_M$ and cannot be assigned to either of $C_1, C_2, D_1, D_2$. Therefore, this constraint will always return \textbf{true} for any valid common bigraph. This thus exhausts all extra constraints in the McSplit MCIS model.

Our original hypothesis that a constraint violation occurs is shown to be a contradiction, and
therefore $S_{MCB}$ must be a valid MCIS solution. This concludes the proof.

\subsubsection{Bound Consistency}

We prove bound consistency by contradiction by setting up the following scenario. Given a MCB instance and solution $(G_1, G_2, G_M)$, our MCIS encoding should return a corresponding solution $M = \{(u_1', v_1'), ..., (u_m', v_m')\}$ where $\sum\limits_{(u', v') \in M} \{\ell(u' ) \neq \textbf{link}\} = |G_M|$. However, let us assume that for some partial solution $K = \{(u_1', v_1'), ..., (u_k', v_k')\}, k < m$ which eventually reaches $M$, the bound check fails and the remaining search tree is pruned, incorrectly preventing $M$ from being found. We set up the bound check so that it is at its strictest, and so we assume that the current best score at this stage is already $|G_M|$ and that $M$ is a valid solution of equal size to the known best maximum.

For the bound to fail, the following must be true: $$|G_M| > |G_K| + \sum\limits_{l \in LC_K} min(|(u \in G_1 \land LC_K(u) = l \ \land \ u \neq (p, i) \in P_{G_1}) |,$$ $$|(v \in G_2 \land LC_K(v) = l \ \land \ v \neq (q, j) \in P_{G_2})|)$$

Where all label classes $l \in LC_K$ contain disjoint subsets of (but do not necessarily together make up the whole sets of) vertices from $G_1$ and $G_2$.

We can infer by our encoding that $|G_K| = \sum\limits_{(u', v') \in K} \{\ell(u) \neq \textbf{link}\}$. We now define $R = \{(u', v') \in \{\{M \setminus K\} \ | \ \ell(u) \neq \textbf{link}\}$, which consists of all non-port element pairs (and therefore all support elements) in $\{M \setminus K\} = \{(u'_{k+1}, v'_{k+1}), ..., (u'_m, v'_m)\}$, the set of assignments which have yet to be added to $K$ in order to reach $M$. By inspection, $|R| = |G_M| - |G_K|$. We substitute this back into the violated bound function as follows: \newpage $$|\{(u', v') \in \{\{M \setminus K\} \ | \ \ell(u) \neq \textbf{link}\}| > \sum\limits_{l \in LC_K} min(|(u \in G_1 \land LC_K(u) = l \ \land \ u \neq (p, i) \in P_{G_1}) |,$$ $$|(v \in G_2 \land LC_K(v) = l \ \land \ v \neq (q, j) \in P_{G_2})|)$$

To simplify, this now states that the number of non-port vertices yet to be assigned for $K$ to reach $M$ must exceed the number of combined pairs of vertices across all non-port label classes which are available for selection. Decoding this back to our original MCB instance, this suggests that adding all remaining compatible support elements to $G_K$ in the pair of compositions $G_1 = C_1 \circ (G_K \otimes id) \circ D_1$ and $G_2 = C_2 \circ (G_K \otimes id) \circ D_2$ is not enough for it to reach a support size of $|G_M|$, and therefore $G_M$ itself cannot be an optimal solution. This contradicts our original hypothesis, and can only occur if $M$ is not an optimal solution to MCIS in the first place. Therefore the bound function will only prune non-optimal solutions, and thus concludes our proof.

\end{document}